\documentclass[sigplan,screen,nonacm]{acmart}
\usepackage{xcolor}
\usepackage{placeins}

\renewcommand\footnotetextcopyrightpermission[1]{}

\newcommand{\dlcb}{DLCB}
\newcommand{\kgen}{kgen}

\begin{document}
\title{\dlcb{}: Ahead-of-Time Compilation for\\ Dynamic Deep Learning}

\newcommand{\nvsc}{\affiliation{%
  \institution{NVIDIA}\city{Santa Clara}\state{CA}\country{USA}}}
\newcommand{\nvrd}{\affiliation{%
  \institution{NVIDIA}\city{Redmond}\state{WA}\country{USA}}}
\newcommand{\nved}{\affiliation{%
  \institution{NVIDIA}\city{Edinburgh}\country{United Kingdom}}}

\author{Alexander Collins}\nved{}\email{acollins@nvidia.com}
\author{Bin Fan}\nvsc{}\email{binf@nvidia.com}
\author{Evghenii Gaburov}\nvsc{}\email{egaburov@nvidia.com}
\author{William Brandon}
\authornote{Work done while at NVIDIA.}
\affiliation{\institution{Independent}\city{Cambridge}\state{MA}\country{USA}}
\email{williambrandon@gmail.com}
\author{Sean Lee}\nvrd{}\email{selee@nvidia.com}
\author{Hanfeng Chen}\nvrd{}\email{hanfengc@nvidia.com}
\author{Vinod Grover}\nvrd{}\email{vgrover@nvidia.com}

\begin{abstract}
  Deep learning workloads are increasingly deployed in settings where tensor
  shapes are not fully known at compile time. Batch sizes vary across requests,
  sequence lengths differ between inputs, and model architectures admit a range
  of spatial resolutions. This dynamism creates a fundamental tension: ahead-of-time
  compiled GPU kernels deliver peak performance but traditionally require
  fully static tensor shapes, while just-in-time compilation supports
  dynamic shapes at the cost of runtime compilation overhead.

  We present \emph{Deep Learning Compiler Backend}, a deep learning compiler that resolves this
  tension through a unified three-tier compilation strategy. For programs with fully static tensor
  shapes, \dlcb{} generates static CUDA C++ kernels ahead of time.  For programs whose tensor rank
  is statically known but whose dimension sizes are dynamic, \dlcb{} generates AOT kernels
  parameterized by those dimensions, compiling once and executing across a range of shapes. For
  programs with tensors of unknown rank, \dlcb{} falls back to JIT compilation at runtime.

  For AOT compilation with dynamic shape tensor, our approach builds a system
  of shape constraints from the input program, and uses a solver to reduce these
  to either a) fixed constraints that produce hard-coded constants in the generated kernel;
  b) unresolved constraints which become kernel parameters passed at launch time, along with
  interpreted host code to check that the constraint holds for a given set of inputs and to compute
  the values to pass.

  Our compiler and runtime are embedded in PyTorch, and uses an automatic shape generalization
  pass to enable AOT compilation with dynamic shapes for supported subsets of the PyTorch programs.

\end{abstract}

\maketitle 

\section{Introduction}

Modern deep learning inference deployments face a pervasive challenge: production workloads rarely
operate on a single fixed set of tensor shapes.  Variable-length sequences in language models,
multi-resolution inputs in vision systems, and dynamically batched requests in serving
infrastructure all mean that the shapes of intermediate tensors change from one invocation to the
next. Yet the GPU kernels that execute these workloads perform best when they are compiled with full
knowledge of the data they will process.

Traditional deep learning compilers resolve this tension in one of two ways.  \emph{Static
compilers} require all tensor shapes to be fixed at compile time.  They deliver good performance:
loop bounds are constants, tile sizes are tuned precisely, and register usage is predictable, but
they must recompile from scratch whenever shapes change, imposing large overhead in scenarios with
dynamic shapes. \emph{Just-in-time compilers} defer compilation to the moment shapes are known,
tolerating dynamic shapes at the cost of compilation latency on the critical path of inference. For
latency-sensitive deployments this overhead is often prohibitive, but can often be mitigated
somewhat by caching compiled binaries.

This paper explores a third path, for the common case in which tensor ranks are statically known,
i.e. the number of dimensions of each tensor is fixed, even if the size along each dimension is not
. It is possible to compile GPU kernels ahead of time that are \emph{parameterized} by dynamic
dimensions. Such kernels are compiled once and execute efficiently across a range of shapes by
receiving dimension sizes as arguments at launch time. The challenge is determining, at compile
time, which dimensions can be treated as constants, which must become parameters, and what host
computation and checks are necessary to ensure the shapes presented at launch are consistent with
the compiled kernel.

\dlcb{} (Deep Learning Compiler Backend) addresses this challenge through a constraint-based
compilation framework. Given a computation graph, \dlcb{} derives a system of \emph{shape
constraints} that capture the relationships between tensor dimensions imposed by each operator. A
constraint solver reduces these constraints to a fixed point, partitioning dimensions into those
that are statically determined (and can be hard-coded) and those that remain free (and become kernel
parameters). Residual constraints (those that cannot be resolved at compile time but must hold at
runtime) are compiled into a lightweight interpreted host program that runs on the CPU before each
kernel launch to verify correctness. The constraints are also used to generate a host code to
compute the concrete dimension values required for the kernels launch parameters.

\dlcb{} supports three compilation paths, selected automatically by the constraint solver: fully
static kernels when all shapes are known; AOT parameterized kernels when ranks are known but
sizes are not, with dynamic dimensions resolved via a lightweight host interpreter at each launch;
and JIT compilation with per-shape caching when ranks themselves are unknown.  The three paths
share a common infrastructure and present a single unified interface.

We integrate \dlcb{} into PyTorch via a bridge layer called \emph{pytorch-dlcb}. PyTorch serves as
frontend: models writen and traced in the usual way, and pytorch-dlcb intercepts the computation
graph before execution. It partitions the graph into subgraphs whose operators are supported by
\dlcb{}, and invokes the \dlcb{} AOT compilation pipeline for each, producing an executable binary
that encapsulates the compiled kernels and their host-side shape verification programs. These
binaries are spliced back into the broader PyTorch execution graph, so that the remainder of the
model runs under PyTorch as normal. Models that mix supported and unsupported operators therefore
benefit from AOT dynamic-shape compilation wherever \dlcb{} applies, while unsupported subgraphs use
standard PyTorch eager execution.

pytorch-dlcb additionally provides an automatic \emph{shape generalization} pass that identifies,
in a statically traced program, which dimensions to abstract into free variables, so the compiled
binary can be reused across shapes without recompilation.

\FloatBarrier
\begin{figure}[t]
  \centering
  \includegraphics[width=\linewidth]{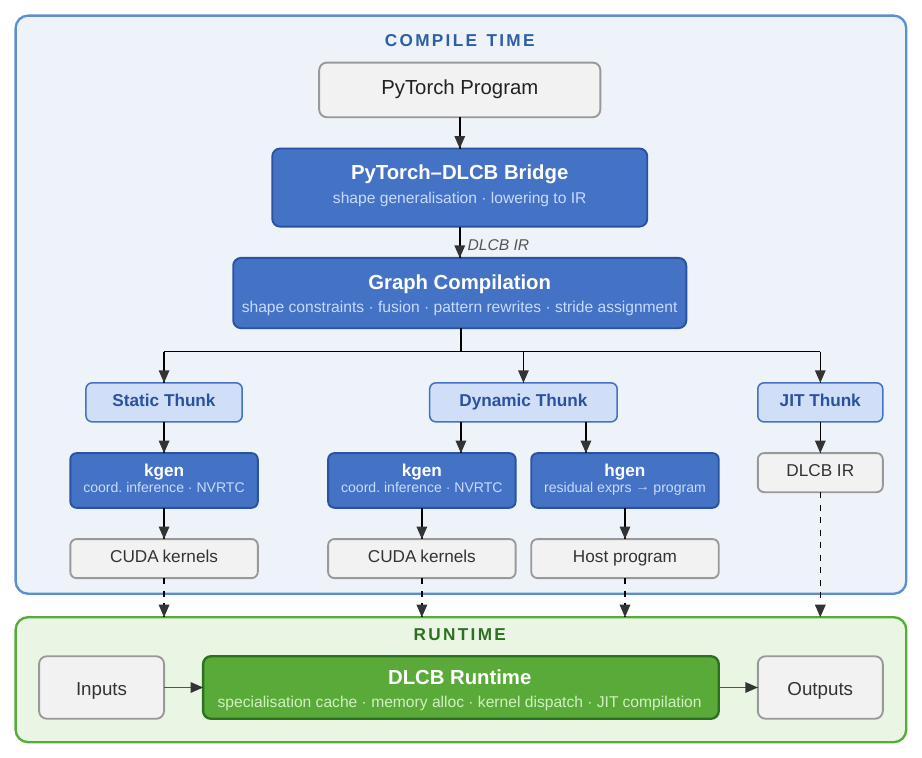}
  \Description{Block diagram of the \dlcb{} compilation pipeline, showing a PyTorch
    program flowing through shape generalization and constraint solving into either
    ahead-of-time generated CUDA kernels or a just-in-time fallback path.}
  \caption{Overview of the \dlcb{} compilation pipeline.}
  \label{fig:system-flow}
\end{figure}

\subsection*{Contributions}

\begin{itemize}

\item \textbf{AOT kernel compilation for dynamic shapes.}
  For computation graphs with statically known tensor rank, we generate CUDA kernels ahead of time
  parameterized by dynamic dimensions.  Shape constraints derived from the operators are reduced by a
  solver to a fixed point, distinguishing hard-coded constants from kernel parameters and runtime
  assertions.

\item \textbf{Compiled host-side shape verification.}  Residual constraints are compiled into a
  compact interpreted program that runs on the CPU before each kernel launch to compute kernel
  arguments and verify shape invariants.

\item \textbf{Automatic JIT fallback.}  When AOT compilation is not possible---typically because
  tensor rank is unknown---the system falls back to JIT compilation, providing a unified interface
  that always produces a runnable artifact.

\item \textbf{Automatic shape generalization.}  pytorch-dlcb automatically identifies which
  dimensions to generalize in a statically traced PyTorch program, producing a dynamic kernel that
  can be reused across different shapes without recompilation.

\end{itemize}

\section{Implementation}

The \dlcb{} compiler and runtime is organised into three stages, summarised in
Figure~\ref{fig:system-flow}.  The PyTorch-\dlcb{} bridge (\S\ref{sec:bridge}) intercepts a
PyTorch program, partitions it into compilable regions, applies shape generalisation, and lowers
each region to the \dlcb{} IR\@.  Graph compilation (\S\ref{sec:graph-compilation}) then takes the
IR, builds a system of shape constraints incrementally as operators are added, runs optimisation
passes (fusion, pattern-matching rewrites, and automatic stride assignment), partitions the graph
into thunks, and produces two compilation artefacts per thunk: a CUDA kernel (or a call into a
vendor library) generated by \kgen{} (\S\ref{sec:codegen}), and a compact host program generated
by the host code generator (\S\ref{sec:hgen}) that resolves dynamic dimensions at launch time.  At
inference time (\S\ref{sec:launch}), the \dlcb{} runtime executes each thunk's host program to
compute kernel arguments and dispatches the corresponding kernel.

\subsection{PyTorch-\dlcb{} bridge}
\label{sec:bridge}

The PyTorch-\dlcb{} bridge is the entry point through which PyTorch programs are lowered to \dlcb{}.
It operates on the PyTorch JIT intermediate representation and is responsible for partitioning the
program into compilable regions, lowering each region to \dlcb{} IR, applying shape generalization,
and registering the compiled binaries as custom operators that are transparent to the rest of the
PyTorch runtime.

\subsubsection{Subgraph identification.}
The bridge maintains a predicate over PyTorch JIT operator names that defines the \emph{supported
subset}: the operators that \dlcb{} can compile.  It walks the JIT graph in topological order,
classifying each node as supported or unsupported.  Consecutive supported nodes are merged into
contiguous \emph{compilation regions}; unsupported nodes form boundaries between regions and are
left in place to be executed natively by the PyTorch runtime.  A node that consumes the output of an
unsupported node contributes a region input; a supported node whose output is consumed by an
unsupported node (or is a graph output) contributes a region output.  The result is a partition of
the JIT graph into an alternating sequence of \dlcb{}-compiled regions and natively executed
operators.

\subsubsection{Lowering to \dlcb{} IR.}
For each compilation region the bridge extracts the subgraph and translates it operator-by-operator
into a \dlcb{} IR graph.  Each tensor input to the region is registered with the \dlcb{} compiler as
either \emph{static} or \emph{dynamic}.  In static mode, the concrete shape and stride values
recorded during tracing are passed directly, allowing \dlcb{} to treat every dimension as a
compile-time constant.  In dynamic mode, fresh symbolic dimension variables are allocated for both
shape and strides, so the compiler treats every dimension as an unknown to be resolved at runtime by
the constraint solver.  Scalar and constant-tensor arguments are always lowered with their concrete
values.  Once lowered, the graph is compiled and the resulting binary is stored in a specialisation
cache keyed on rank and element type (see \S\ref{sec:launch}); subsequent invocations with the same
rank reuse the cached binary without recompilation.

\subsubsection{Shape generalization.}
When a model is traced under static shapes, the bridge applies an automatic \emph{shape
generalization} pass before lowering.  The pass takes the JIT IR and automatically identifies which
dimensions are safe to abstract into free symbolic variables, producing a parameterized kernel that
can be reused whenever the program is called with different concrete shapes, eliminating
recompilation and removing the burden of manual shape annotation from the user.  The pass proceeds
in four steps.

\paragraph{Identifying shape consumers.}
A predicate determines which argument positions of which operators can safely receive a dynamic
(symbolic) shape value.  For example, the second argument of \texttt{aten::reshape}---the target
shape list---is a shape consumer: replacing its concrete integer values with symbolic variables is
safe because the operator is defined entirely by the shape it is given.  The predicate covers all
such argument positions across the supported operator set.

\paragraph{Tracing dim-computing subgraphs.}
For each shape-consuming use, the pass recursively walks the def-use chain of the shape argument to
collect all nodes that compute it.  Recognised node kinds are:
\begin{itemize}
  \item \emph{Leaf sources}: \texttt{aten::size} and constant nodes.
  \item \emph{List construction}: list-construction nodes, whose inputs are
    themselves recursively traced.
  \item \emph{Arithmetic on dims}: \texttt{aten::add}, \texttt{aten::sub},
    \texttt{aten::rsub}, \texttt{aten::mul}, \texttt{aten::floordiv}, and
    truncating \texttt{aten::div}.
  \item \emph{Unary and coercion nodes}: \texttt{aten::neg} and scalar
    coercion operations.
\end{itemize}
The result is a topologically ordered list of nodes forming the shape-computation subgraph rooted at
the consumer's input.

\paragraph{Cloning and redirecting.}
The pass iterates over every node in the JIT block.  For each shape-consuming use it collects the
corresponding dimension-source cluster, clones every node in the cluster into fresh nodes inserted
immediately before the consumer, and redirects the consumer's input to the cloned output.

Cloning is necessary because a shape-computing node---e.g.\ an \texttt{aten::size} call---may be
shared between a static use and a shape-consuming use.  After cloning, dead-code elimination removes
the originals that are no longer needed, and the cloned dimension-computing nodes are pulled into
the \dlcb{} compilation group, where they are compiled symbolically.

\paragraph{Symbolic tensor inputs.}
When a tensor input to a compilation group is registered, the bridge checks whether it is static or
dynamic.  For a static input, concrete integer lists for the shape and strides are passed directly
to the compiler.  For a dynamic input, fresh symbolic dimension variables are allocated for both
shape and strides, so that \dlcb{} treats every dimension as an unknown.  When the
layout-preservation option is enabled, the strides are sorted to recover the memory layout order
(e.g.\ row-major or channel-last), and this structural constraint is encoded alongside the symbolic
dimensions, allowing \dlcb{} to generate layout-aware kernels while keeping dimension sizes
symbolic.

\subsubsection{Broadcast optimization.}
PyTorch traces record concrete shapes, which means the bridge knows---before any lowering
occurs---which dimensions have size~1 in the traced execution and are therefore being broadcast.
The bridge exploits this information as a \emph{broadcast hint}: for each binary operator whose
inputs differ in shape in the trace, the size-1 dimensions are annotated as statically broadcast.
\dlcb{} uses these annotations during code generation to emit specialised broadcast paths that avoid
redundant index arithmetic for the broadcast dimension, producing more efficient kernels than would
be possible if every dimension were treated as fully general.  Because the hint is derived from the
static trace rather than the dynamic inputs, it is validated at runtime: the host program includes a
check that verifies the broadcast dimension is indeed~1 before dispatching the specialised kernel,
falling back to the general path if the check fails.

\subsection{Graph compilation}
\label{sec:graph-compilation}

Once a subgraph has been extracted, operators are added to the \dlcb{} IR graph one at a time.
As each operator is added, it contributes equalities to a system of \emph{shape constraints} that
capture the relationships between tensor dimensions imposed by that operator: a pointwise binary
operator asserts broadcast compatibility between its two inputs and equates the output shape to the
broadcast result; a matrix multiplication constrains the inner dimensions of its operands to match
and defines the output shape from the outer dimensions; a reshape operator equates the product of
input dimensions to the product of output dimensions; and so on, covering reductions,
gather/scatter, and other supported operation classes.  Shape inference is run incrementally after
each operator is added, propagating the new constraints through the solver immediately.  This
provides fail-fast behaviour: if an operator introduces a contradiction---for example, a shape
mismatch between two operands---the error is detected at graph-construction time rather than after
the full graph has been built, giving immediate feedback on the source of the error.

\subsubsection{Optimization Passes}
\label{sec:optimizations}
Once the graph is fully constructed, a series of optimization passes are applied before code
generation.  Automatic stride assignment initialises any tensor whose layout has not been explicitly
specified to row-major order.  Fusion passes merge compatible operator sequences, including
pointwise fusion and folding transposes into adjacent matrix multiplications, reducing kernel
launches and improving memory-access locality.  Pattern-matching rewrites identify common operator
subgraphs and substitute more efficient implementations; for example, the canonical layer
normalisation subgraph---a mean subtraction followed by variance computation and division---is
recognised and replaced by a single fused layernorm operator, eliminating redundant mean evaluations
and enabling a hardware-optimised implementation.

\subsubsection{Partitioning}
\label{sec:partitioner}

After the optimization passes, the \dlcb{} IR graph is partitioned into \emph{thunks}.  A thunk is a
general-purpose compilation unit: it accepts a fixed set of tensor inputs, produces a fixed set of
tensor results, and encapsulates all the computation needed to do so.  The partitioner inspects each
operator and groups operators into thunks according to how they will be compiled and executed.

\emph{Library operators}---principally matrix multiplications and convolutions---each form their own
singleton thunk backed by a vendor-optimised library call.  Matrix multiplications are dispatched to
cuBLAS or cuBLASLt, and convolutions to cuDNN\@.  The partitioner does not attempt to fuse these
operators with their neighbours, since the vendor libraries apply their own internal optimisations.

All remaining operators---pointwise and elementwise operations (unary and binary arithmetic,
activations, type casts), reductions, and index transformations such as reshapes and
transposes---are grouped into thunks backed by custom CUDA C++ kernels generated by \kgen{}
(\S\ref{sec:codegen}).  Consecutive such operators are merged into a single thunk wherever
possible, allowing \kgen{} to emit one fused kernel covering the entire sequence and eliminating
intermediate memory traffic between operators.

\subsubsection{Kernel code generation (\kgen{})}
\label{sec:codegen}

Given the fused operator graph and the dimension assignments produced by the constraint solver,
\kgen{} uses \emph{coordinate-based inference} to determine, for each output element, which input
coordinates are required.  Starting from the set of output tensors and their coordinate spaces,
\kgen{} propagates coordinates backwards through each operator in topological order, building a map
from output coordinates to the input coordinates and intermediate values needed to compute them.  It
then emits a CUDA kernel that iterates over output coordinates, evaluating this map to read inputs
and write outputs.

Dynamic dimension values---those left as free parameters by the constraint solver---appear in the
generated kernel as integer parameters.  The values of these arguments are computed at launch time
by the host code interpreter described in \S\ref{sec:hgen}, ensuring that a single compiled kernel
binary executes correctly across any shapes for which the rank and layout structure match.

More details of how \kgen{} generates kernels parameterized by dynamic dimensions are described in
\S\ref{sec:dynkgen}.

\subsubsection{Host code generation}
\label{sec:hgen}

After constraint solving, some dimension expressions remain unsolved: they depend on runtime values
that are not yet known.  The host code generator compiles these residual expressions into a compact
interpreted program that runs on the CPU immediately before each kernel launch---i.e. before a given
thunk is launched.  It produces two artefacts: an instruction sequence and a context record (a table
mapping tensor identifiers and operator ports to the host program's input and output slots).
Together they allow the kernel launcher to supply runtime shapes, verify shape constraints, compute
derived dimensions, and pass the results to the kernel as launch arguments.

\paragraph{Input variable allocation.}
First, the generator scans all input tensors in the thunk.  For each input tensor whose rank is
statically known, it inspects every shape dimension.  Dimensions already resolved to compile-time
constants require no runtime value and are assigned no input slot.  Each remaining non-literal shape
dimension is assigned a sequential input variable index, and the mapping is stored in a per-tensor
input table.  The same variable index is then introduced into the constraint system as a named
literal, effectively naming the runtime value of that dimension so that inference can propagate it
through the constraint graph.

\paragraph{Output slot registration.}
The generator then constructs an ordered list of output expressions, one slot per non-literal
dimension value that the kernel will need at launch time.  Three categories of output are
registered:

\begin{itemize}
  \item \emph{Tensor shapes and strides.}  For every input, output, and intermediate tensor with a
    known rank, each non-literal shape and stride dimension is assigned a sequential output slot.
    Statically resolved dimensions are assigned no slot.
  \item \emph{Kernel generator constants.}  Some graph partitions are compiled to an opaque
    thunk---such as for cuDNN and cuBLAS ops.  Such thunks may declare expressions for constants
    that need to be dynamically computed.  Each is assigned an output slot so that it is computed by
    the host program.
  \item \emph{Dynamic operator attributes.}  Operators that carry dimension-list attributes
    (e.g.\ reshape target shapes) expose those as outputs, one slot per attribute element.
\end{itemize}

\paragraph{Inference and stride promotion.}
With input variables named and output expressions registered, the generator re-runs the constraint
solver on the combined constraint set which consists of the remaining constraints for the graph,
plus all the constraints added by input variable allocation and output slot registration.

If a stride output remains unsolved after inference---its expression still contains a free
variable---the generator promotes that stride to an additional input variable.  It inserts a new
named-literal equality constraint equating the stride expression to the fresh input variable and
re-runs inference.  This handles operators whose output strides cannot be derived algebraically from
shape information alone and must therefore be read from the tensor descriptor at launch time.

\paragraph{Program representation.}
The host program is a flat, ordered sequence of instructions executed by a
simple register machine with three classes of storage: \emph{inputs}
(read-only integer values supplied at runtime), \emph{outputs} (writeable
slots read by the kernel launcher after execution), and \emph{registers}
(temporaries private to the program).  The instruction set comprises sixteen
opcodes: data movement (\texttt{CP}, \texttt{LD}, \texttt{ST},
\texttt{MOV}); arithmetic on two source registers
(\texttt{ADD}, \texttt{SUB}, \texttt{MUL}, \texttt{DIV},
\texttt{DIV\_CEILING}, \texttt{MOD}, \texttt{MIN}, \texttt{MAX});
\texttt{BROADCAST}, which applies the NumPy broadcasting rule and
returns \texttt{false} on incompatible shapes; \texttt{CHK\_EQ}, which
compares two registers and returns \texttt{false} if they differ; and
\texttt{FASTINTOP\_CONSTS}, which writes the precomputed magic
constants $(M, s, n)$ for a divisor read from one source register.

\paragraph{Compilation.}
The host program compiler translates the ordered output list and the residual
constraint set into the instruction sequence by recursive descent with
memoisation.  For each output slot the compiler inspects the expression
rooted at that slot:
\begin{itemize}
  \item If the expression is a named literal (a direct alias for an input
    variable), it emits a single \texttt{CP} instruction.
  \item Otherwise it recursively compiles the expression tree, which returns
    the register holding the computed value, and emits \texttt{ST} to write
    that register to the output slot.
\end{itemize}
For each residual constraint whose both sides are fully resolved, the
compiler compiles both sides and emits \texttt{CHK\_EQ}.

The recursive compilation dispatches on the expression kind: integer literals emit \texttt{MOV};
named literals emit \texttt{LD} from the corresponding input slot; arithmetic expressions compile
their operands and emit the matching instruction; and fast-integer-division constants compile the
divisor and emit \texttt{FASTINTOP\_CONSTS}, returning the register for the requested
$(M, s, n)$ component.  Results are memoised by expression identity.

\paragraph{Storage and invocation.}
The compiled instruction sequence and the context record are stored as part of the thunk.  At
kernel launch time, the \dlcb{} runtime fills the input vector from the concrete shapes and strides
of the live input tensors via the context record, then runs the interpreter.  The interpreter
maintains a flat register file and dispatches each instruction in order via a switch statement;
\texttt{BROADCAST} and \texttt{CHK\_EQ} can return \texttt{false} early to signal an incompatible
shape or violated constraint, while all other instructions are infallible.  On success the output
vector contains the resolved shapes, strides, and kernel constants, which are passed directly to
the CUDA kernel as launch arguments.  The host program is compiled once at graph compilation time
and interpreted cheaply on every invocation, keeping symbolic computation off the inference
critical path.

\subsection{Runtime kernel launch}
\label{sec:launch}

At inference time, pytorch-dlcb maintains a per-subgraph cache of compiled thunks.  On each
invocation it computes a \emph{specialisation key} from the live input tensors---encoding element
type, rank, and (in static mode) concrete shape and stride values, plus the full content of any
constant tensor or scalar arguments---and looks up the key in the cache.  On a hit the cached thunk
is executed immediately.  On a miss the \dlcb{} compiler pipeline is invoked to compile the
subgraph, a runtime context is created with bound cuBLAS, cuBLASLt and cuDNN handles, GPU
buffers for embedded data constants are allocated and initialised, and the resulting thunk is
inserted into the cache before execution.  Three kinds of thunk are produced, depending on what was
known about tensor shapes at compile time.

\paragraph{Static thunks.}
When all tensor shapes are fully known at trace time, the subgraph is compiled into a
\emph{static thunk}.  The specialisation key includes the concrete shape and stride values of
every input, so a distinct thunk exists for each combination of shapes encountered at runtime.
The thunk contains fully static CUDA kernels whose launch configurations---grid dimensions, block
dimensions, and shared memory size---are resolved entirely at compile time.  At invocation the
runtime proceeds directly to memory allocation and kernel dispatch; there is no host-side shape
computation.

\paragraph{Dynamic (AOT) thunks.}
When tensor ranks are statically known but dimension sizes are not, the subgraph is compiled into
a \emph{dynamic thunk}.  The specialisation key includes only rank and element type, so a single
thunk handles all inputs with the same rank regardless of their concrete sizes.  At invocation the
runtime runs the thunk's host program (§\ref{sec:hgen}) to resolve the dynamic dimensions, then
proceeds as follows:

\begin{enumerate}
  \item \textbf{Shape registration.}  The concrete sizes and strides of each live input tensor are
    recorded.
  \item \textbf{Shape commitment.}  The host program interpreter is run, propagating the
    registered shapes through the compiled constraint program to resolve output shapes, strides,
    and other necessary constants.
  \item \textbf{Scratch space computation.}  Each thunk wrapper is queried for the temporary GPU
    memory it requires given the now-known shapes.
  \item \textbf{Memory allocation and kernel dispatch.}  A single temporary tensor pool is
    allocated.  For each thunk wrapper, input tensors are mapped from the live PyTorch tensors,
    output tensors are allocated from the shapes produced by the host program, and intermediate
    tensors are sub-allocated from the pool and freed by reference counting as soon as their last
    consumer completes.  Each wrapper is then dispatched with the resolved storage pointers and
    the current CUDA stream.
\end{enumerate}

\paragraph{JIT thunks.}
When tensor rank is not statically known, the constraint solver cannot produce a parameterized
kernel ahead of time.  The subgraph is instead stored as a \emph{JIT thunk} containing the
uncompiled \dlcb{} IR\@.  On the first invocation with a given set of concrete shapes the IR is
compiled on-demand, the resulting kernels are cached inside the thunk, and execution proceeds.
Subsequent invocations with the same shapes bypass compilation entirely.  From pytorch-dlcb's
perspective the JIT path is transparent: the same cache-lookup mechanism is used, and the outer
specialisation cache avoids redundant compilation by keying on rank rather than concrete values.

\section{Constraint Solving}
\label{sec:constraints}

Shape constraints are solved by a generic fixed-point inference engine
parameterised over the graph representation it operates on.  The engine
is instantiated for two distinct purposes: solving dimension-size
constraints (the \emph{dimlist inference} described here) and solving
coordinate-set constraints used during kernel code generation (\S\ref{sec:coord-inference}).

\paragraph{Constraint representation.}
Every constraint is an equality between two \emph{dimension expressions},
represented by nodes in a DAG.  Six node kinds are used:

\begin{itemize}
  \item \emph{Integer literals} --- a concrete integer constant.
  \item \emph{Named literals} --- a named constant whose numerical value is
    unknown at compile time but will be supplied at runtime (e.g.\ an input
    tensor's batch dimension).  Named literals are preserved by the solver
    and appear in the generated host program as inputs.
  \item \emph{Free variables} --- unknowns to be eliminated by the solver.
  \item \emph{Arithmetic expressions} --- a binary arithmetic expression with
    operator drawn from \{\textsc{add}, \textsc{sub}, \textsc{mul},
    \textsc{div}, \textsc{div\_ceiling}, \textsc{mod},
    \textsc{min}, \textsc{max}, \textsc{scale}, \textsc{broadcast}\}.
  \item \emph{Shape lists} --- an ordered list of dimension expressions,
    representing a full tensor shape.
  \item \emph{List operations} --- a list operation with operator drawn from
    \{\textsc{append}, \textsc{broadcast}\}, used to express broadcast and
    concatenated-rank shapes.
\end{itemize}

Each node tracks its set of free variables and a content-addressed hash,
enabling $O(1)$ structural equality tests.  The constraint set is a list of
equality pairs $(l, r)$, cross-indexed by vertex so that all constraints
mentioning a given node can be found in $O(\log n)$.  Reference counting on
nodes supports prompt dead-code elimination when a node is no longer
referenced by any constraint or graph root.

\paragraph{Fixed-point algorithm.}
The solver iterates over the constraint set, applying the first applicable
rewrite rule at each step and restarting the scan whenever a rule fires,
until no rule applies.  The rules, in priority order, are:

\begin{enumerate}

  \item \textbf{Trivial equality.}
    $e = e \;\Rightarrow\; \varnothing$.
    Remove the constraint and prune the vertex.

  \item \textbf{Variable substitution.}
    $x = e$ where $x \notin \mathit{FV}(e)
    \;\Rightarrow\;$ substitute $e$ for $x$ everywhere in the graph,
    then erase the constraint.
    This is the primary elimination step: it removes the variable $x$
    from the problem entirely.

  \item \textbf{Constant conflict check.}
    $c_1 = c_2$ where both sides are integer literals
    $\;\Rightarrow\;$ fail if $c_1 \neq c_2$.
    Contradictory constraints cause the solver to return failure.

  \item \textbf{Batch list substitution.}
    $[x_1, \ldots, x_n] = [c_1, \ldots, c_n]$ where all
    $x_i$ are variables and all $c_i$ are literals
    $\;\Rightarrow\;$ substitute each $c_i$ for $x_i$ simultaneously.
    This optimised rule avoids the overhead of unpacking the list into
    $n$ separate constraints when the solution is immediate.

  \item \textbf{Reordering.}
    $e = x$ $\;\Rightarrow\;$ flip to $x = e$.
    Variables are normalised to the left-hand side so that rule~2
    can apply.

  \item \textbf{List unpacking.}
    $[x_1, \ldots, x_n] = [y_1, \ldots, y_n]
    \;\Rightarrow\; x_1 = y_1,\; \ldots,\; x_n = y_n$.
    A rank mismatch produces a failure.

  \item \textbf{Append unpacking.}
    Two rules strip one element at a time from the trailing or leading
    end of an \textsc{append} expression equated with a concrete list,
    generating one scalar equality per element.  For example,
    $e_1 \mathbin{@} [x_1, \ldots, x_n] = [y_1, \ldots, y_n]$
    produces $[x_1, \ldots, x_n] = [y_1, \ldots, y_n]$ and
    $e_1 = [\,]$.

  \item \textbf{Arithmetic inversion.}
    Three rules solve simple linear equations when one side is a
    concrete literal:
    \begin{align*}
      e \cdot c_1 &= c_2 \;\Rightarrow\; e = c_2 / c_1
        \quad (\text{when } c_1 \mid c_2), \\
      e + c_1 &= c_2 \;\Rightarrow\; e = c_2 - c_1, \\
      e - c_1 &= c_2 \;\Rightarrow\; e = c_1 + c_2.
    \end{align*}

  \item \textbf{Common-subexpression cancellation.}
    $(f\; a\; b) = (f\; a\; c) \;\Rightarrow\; b = c$
    for additive and subtractive operators with a shared operand.

  \item \textbf{Partial-expression substitution.}
    $(f\; e_1 \ldots e_k) = c$ where the expression node has remaining
    uses in the graph $\;\Rightarrow\;$ substitute $c$ for the expression
    node throughout the graph while keeping the constraint alive.  This
    propagates known values to all sites that compute the same
    sub-expression.

\end{enumerate}

After every substitution the graph is updated: affected ancestor nodes
have their hashes and free-variable sets recomputed in topological order,
and any node whose reference count drops to zero is pruned by a depth-first
sweep.

\paragraph{Sufficiency in practice.}
Although the rule set does not handle arbitrary systems of equations, we have found it sufficient
for all constraints arising from real-world programs.  The key reason is that constraints are not
generated arbitrarily: they are produced mechanically by a fixed set of supported operators, each of
which contributes a small, well-understood family of equalities---broadcast compatibility, reshape
products, matmul inner-dimension matching, and so on.  The structural regularity of this constraint
vocabulary means that the rules above cover every pattern that can actually appear.  General-purpose
constraint solvers (e.g.\ SMT-based approaches) would be capable of handling a much wider class of
constraints but would impose unnecessary overhead and complexity; the bounded operator set makes
them unnecessary.

\paragraph{Expression simplification.}
Before a node is used in a new constraint or substituted into the graph, it is passed through a
simplification pass.  Simplification performs constant folding for all arithmetic operations, and
eliminates algebraic identities such as $x + 0 \Rightarrow x$, $x \times 1 \Rightarrow x$, $x / 1
\Rightarrow x$, and $\min(x, x) \Rightarrow x$.  Nodes are content-addressed, so structurally
identical expressions are automatically deduplicated.

\paragraph{Outcome.}
After the solver reaches a fixed point, three kinds of dimension
expression remain in the constraint graph:
\begin{itemize}
  \item An \emph{integer literal} --- the dimension is fully determined at
    compile time and can be hard-coded in the generated kernel.
  \item A \emph{named literal} --- the dimension is a direct function of a
    runtime input (e.g.\ a linear expression of the input batch size).  The
    named literal is recorded in the kernel parameter table as an argument to
    be supplied at launch time; statically resolved dimensions receive no
    entry.
  \item A \emph{residual expression} containing free variables --- the
    dimension depends on shapes that cannot be determined even symbolically.
    Residual constraints are compiled by the host code generator
    (\S\ref{sec:hgen}) into a lightweight interpreted program that evaluates
    them at kernel launch time.
\end{itemize}

\section{Dynamic Kernel Generation}
\label{sec:dynkgen}

\kgen{} is the component of \dlcb{} responsible for generating CUDA kernels for pointwise thunks
(\S\ref{sec:partitioner}).  Its central challenge is that a pointwise thunk can contain arbitrary
compositions of operators---elementwise arithmetic, reshapes, broadcasts, reductions, gathers, and
index transformations---each of which introduces different index arithmetic relating output
positions to input positions.  Rather than implementing this arithmetic separately for each
operator combination, \kgen{} derives it automatically via \emph{coordinate inference}: a backward
propagation algorithm that, given abstract coordinates assigned to the output tensors, computes the
corresponding input coordinates needed at every node in the graph.  The generated CUDA kernel
iterates over output coordinates, evaluates the inferred map to load the required inputs, and
writes results to output memory.

\subsection{Coordinate Representation}

A \emph{coordinate} in \kgen{} is a logical position in a tensor's index space, represented as an
expression over free variables and dimension values.  The representation has two levels.

At the set level, a \emph{coordinate set} represents the full multi-dimensional position of a
tensor element.  Coordinate sets are either free variables (symbolic positions not yet resolved),
literal sets (concrete index tuples), or \emph{derived sets} produced by a coordinate-set operator
such as \textsc{reshape}, \textsc{broadcast}, \textsc{transpose}, or \textsc{gather}.

At the expression level, each scalar dimension within a coordinate set is a \emph{coordinate
expression}: an arithmetic expression built from integer literals, free variables, references to
dynamic tensor dimensions, and binary operators $\{+, -, \times, \div, \bmod, \min, \max\}$.
Coordinate expressions are content-addressed nodes in a DAG, so structurally identical
sub-expressions are automatically shared and computed only once in the generated kernel.

A \emph{coordinate space} wraps a coordinate set with optional \emph{predicate} bounds that define
when the coordinate is valid---for example, when a thread covers a partial tile at a tensor
boundary.  Scalar coordinate spaces hold a single element position; vector coordinate spaces bundle
a small fixed number of scalar positions for SIMD access patterns.

\subsection{Output Annotation and the Inference Problem}

\kgen{} begins by annotating each output tensor of the thunk with a fresh free-variable coordinate
set.  This variable represents the logical position of whichever output element a given CUDA thread
is responsible for computing; the kernel is launched with a thread grid that covers all positions
in the output coordinate space, so every output element is assigned to exactly one thread.

The coordinate inference problem is then: for each operator in the graph, and for each output
coordinate assigned to that operator's result, determine which coordinate of each input is required
to compute it.  Solving this problem for every operator in topological order---from outputs back to
inputs---yields a complete coordinate map that can be evaluated independently at each thread.

\subsection{Coordinate Inference}
\label{sec:coord-inference}

Coordinate inference is a backward propagation pass over the operator graph.  For each operator, a
constraint generator produces a \emph{reverse coordinate transformation}: a function from the
output coordinate to the input coordinate(s) needed to compute it.

\paragraph{Elementwise operators.}
For unary and binary pointwise operators the reverse transform is the identity: input and output
coordinates are identical.  For a binary operator with broadcast semantics, the transform applies
\textsc{broadcast\_rev}: if output dimension $d$ has size greater than~1 but the corresponding
input dimension has size~1, the input coordinate for that dimension is mapped to~0.

\paragraph{Reshape.}
A reshape from shape $(d_0, d_1, \ldots)$ to shape $(e_0, e_1, \ldots)$ maps output coordinates
to input coordinates via integer division and modulo.  Given a linearised output index $x$, the
reverse transform recovers the input multi-index as
$\bigl(\lfloor x / s_1 \rfloor \bmod d_0,\; \lfloor x / s_2 \rfloor \bmod d_1,\; \ldots\bigr)$
where $s_k$ are the output strides.  These expressions appear inline in the generated kernel; when
divisors are dynamic dimensions the fast-integer-division constants supplied by the host program
(\S\ref{sec:hgen}) are substituted to avoid hardware divide instructions on every thread.

\paragraph{Reduction.}
A reduction over axis $k$ introduces multiple input coordinates per output coordinate: given output
coordinate $(i, \ldots)$ with axis $k$ absent, the reverse transform produces input coordinates
$(i, \ldots, j, \ldots)$ for all $j \in [0, d_k)$.  The generated kernel wraps the inner
computation in a loop over $j$, accumulating the partial result.

\paragraph{Gather.}
A gather operator reads data at positions specified by an index tensor.  Its reverse transform
produces two sets of input coordinates: the output coordinate applied to the index tensor to
retrieve the gather index, and the retrieved index value applied to the data tensor, introducing a
data-dependent memory access in the generated code.

\paragraph{Fixed point.}
The inference engine iterates over the constraint set, applying rewrite rules that substitute known
coordinate expressions for free variables until no rule applies.  The rule structure mirrors the
shape constraint solver (\S\ref{sec:constraints}).  At the fixed point every node in the graph has
a resolved coordinate expression relating it to the output coordinate variables.

\subsection{CUDA C++ Code Generation}

Once the coordinate map is complete, \kgen{} generates a CUDA C++ kernel structured as follows.  The
generated CUDA C++ source is compiled to a binary using NVIDIA's NVRTC runtime compilation library.
The resulting cubin is extracted and stored in the thunk alongside the host program.  NVRTC runs
once at graph compile time; at inference time the cubin is loaded by the CUDA driver and launched
without further compilation overhead.

\paragraph{Thread-to-coordinate mapping.}
The kernel is launched with a one- or two-dimensional thread grid that covers all positions in the
output coordinate space.  Each thread computes its output coordinate from \texttt{blockIdx} and
\texttt{threadIdx} using the output tensor's strides.

\paragraph{Dynamic dimension parameters.}
Any dimension value not resolved to a compile-time constant appears as an entry in a
\texttt{const int32\_t*} buffer passed to the kernel as a parameter.  Each such value is loaded
once at the start of the kernel into a local variable; all index arithmetic references the local
variable.  The host program (\S\ref{sec:hgen}) fills this buffer before each launch.

\paragraph{Index arithmetic and memory access.}
For each input tensor the kernel evaluates the inferred coordinate expression to compute a flat
memory index and loads the value.  Scalar coordinate expressions are emitted as inline C++
arithmetic; vector coordinate spaces produce vectorised loads (e.g.\ \texttt{float4}) where
alignment permits.  Predicate conditions arising from boundary tiles or masked reductions are
emitted as guards, substituting a default value (zero or the reduction identity) when the
predicate is false.

\paragraph{Operator evaluation.}
Operators are emitted in topological order, reusing values shared across multiple consumers.  Each
operator maps to an inline C++ expression or intrinsic.  The final result is written to the output
tensor at the thread's output coordinate.

\subsection{Dynamic Vectorization}
\label{sec:dynvec}

GPU memory bandwidth is best exploited when threads issue wide, aligned loads and stores using CUDA
vector types such as \texttt{float4} or \texttt{half8}.  For kernels with statically known shapes,
the required conditions---that the last dimension is contiguous in memory and its size is divisible
by the vector width---can be verified at compile time, so a single fixed-width kernel is emitted.
Dynamic-shape kernels do not have this luxury: dimension sizes and memory strides are not known
until the kernel is launched.

\paragraph{Compile-time width selection.}
\kgen{} selects a candidate vector width $v$ at compile time using two inputs: (i) the
hard-coded maximum width of an 8-element wide vector, and (ii) the element type of the
tensors involved, which constrains which CUDA vector types are valid (e.g.\ \texttt{float32} only
supports widths of~1, 2, and~4).  If any operator in the DAG uses a type for which no vector
load exists---such as complex types---the candidate width is reduced to~1 and a purely scalar kernel
is emitted.  Subject to these constraints, the code generator targets the largest width $v$ that
is valid.

\paragraph{Runtime guard.}
Because divisibility and contiguity cannot be established at compile time, the generated kernel
contains both a vectorized branch and a scalar fallback branch, selected by a runtime boolean guard.
\kgen{} always attempts to vectorize the innermost (last) dimension.  The guard evaluates two
conditions against the dynamic dimension buffer loaded at kernel entry:
\begin{enumerate}
  \item \emph{Divisibility}: the last-dimension size of every input and output tensor is divisible
    by~$v$.
  \item \emph{Contiguity}: the stride of the last dimension of every input and output tensor
    equals~1 (i.e.\ elements are contiguous in memory).
\end{enumerate}
If both conditions hold the vectorized branch executes; otherwise the scalar fallback ($v=1$)
executes.

\paragraph{Coordinate grouping and code generation.}
When the vectorized branch is taken, \kgen{} generates a kernel in which the thread count is divided
by $v$: each thread is responsible for $v$ consecutive elements along the vectorized dimension.
After coordinate inference resolves the symbolic index map, the code generator identifies groups of
$v$ consecutive coordinate expressions---using the algebraic simplification rules of the constraint
engine (\S\ref{sec:constraints}) to establish consecutiveness---and merges each group into a single
\texttt{vector} coordinate expression.  These vector coordinates expressions are lowered to CUDA
vector loads and stores, maximizing memory-access efficiency.

\section{Evaluation}
\label{sec:eval}

%
%
%
%

We evaluate \dlcb{} on the Torchvision benchmark suite, which contains a wide range of image
classification and segmentation models.
All experiments are run on an NVIDIA H100 GPU with CUDA~13.2, cuDNN~9.20, and PyTorch~2.11. All
models were run using half precision (fp16).  Each model is run for 3 warmup iterations followed
by 5 timed iterations.




\subsection{End-to-end inference throughput}

\begin{figure*}[p]
  \centering
  \setlength{\abovecaptionskip}{4pt}
  \vspace*{\fill}
  \includegraphics[scale=1]{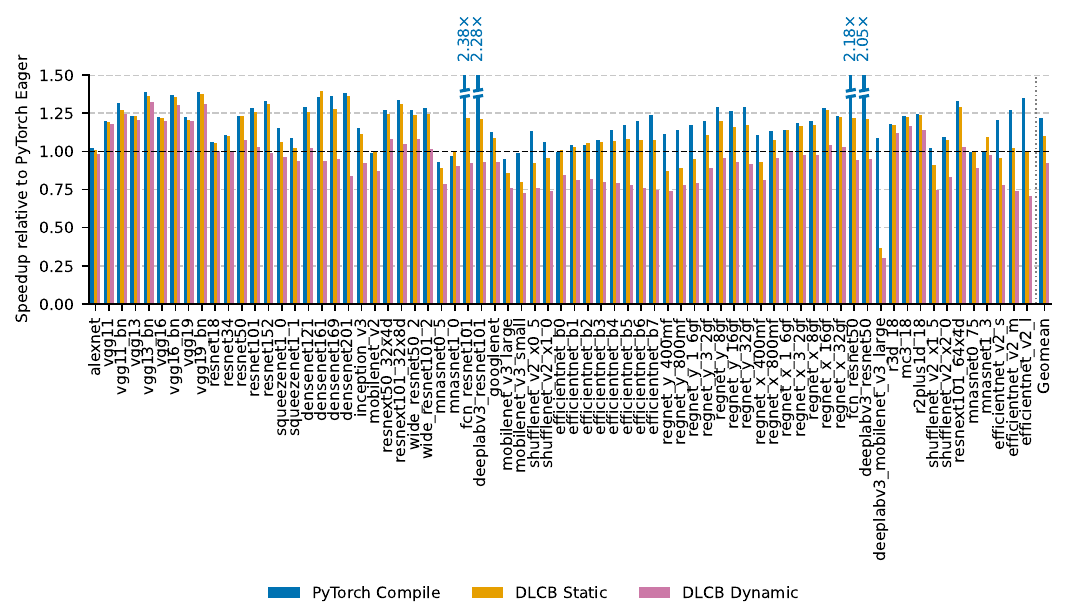}
  \Description{Bar chart of speedup over PyTorch eager execution for each TorchVision
    classification model, grouped by compilation strategy, with a geometric mean bar
    for each strategy.}
  \caption{Speedup over PyTorch eager on TorchVision classification models. Higher is
    better. Geometric mean is shown for each compilation strategy.}
  \label{fig:torchvision}

  \vspace*{\fill}

  \includegraphics[scale=1]{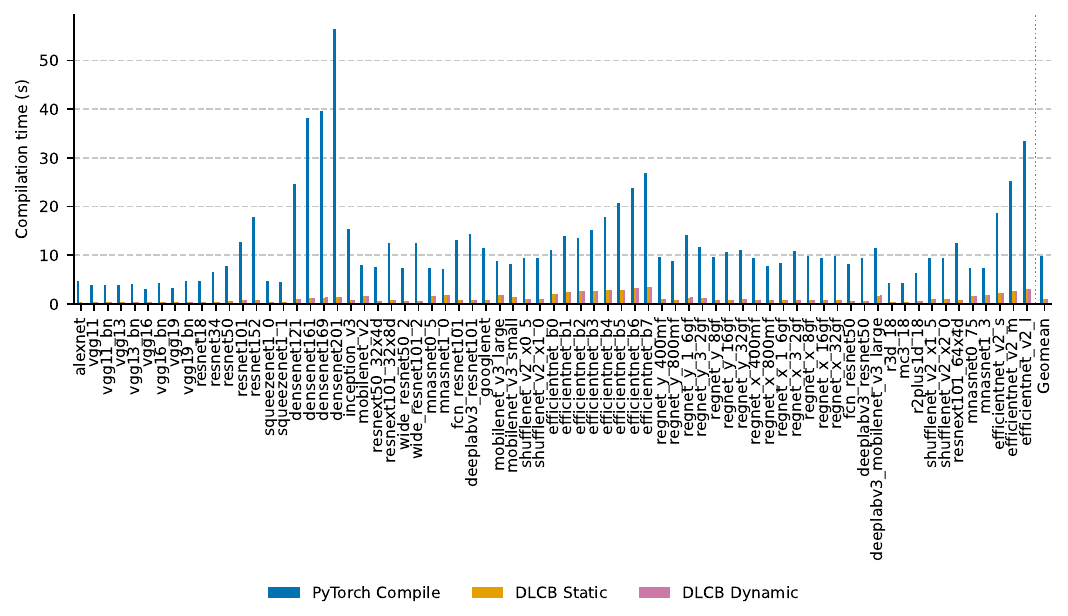}
  \Description{Bar chart of compilation time in seconds for each TorchVision model,
    comparing the PyTorch compiler against \dlcb{}-static and \dlcb{}-dynamic.}
  \caption{Compilation time on TorchVision models.  PyTorch compiler and \dlcb{}-static require one
    compilation per distinct shape encountered; \dlcb{}-dynamic requires exactly one compilation
    regardless of how many shapes are seen at runtime.}
  \label{fig:compile-time}
  \vspace*{\fill}
\end{figure*}

Figure~\ref{fig:torchvision} shows speedup over PyTorch eager on the TorchVision classification
suite.

Overall, the static mode achieves a 1.10x speedup and the dynamic mode 0.92x speedup relative to eager. Since this paper focuses on dynamic-shape compilation, we discuss the dynamic mode in more detail. Dynamic mode shares all optimisation passes with static mode, with several passes extended to handle symbolic shapes (e.g., the pattern matcher matches symbolic dimensions, and the kernel generator guards vectorised code with runtime checks). As a result, dynamic mode beats eager on some models, but on average is slower than eager due to three issues that are difficult to resolve fully.

\textbf{Constant folding and CSE limitations.} Because shapes and strides are not constants during kernel generation, expressions that depend on them cannot be constant-folded or CSE'd. All address computations must therefore use integer division or modulo, which is expensive on the GPU; comparing kernels generated from the same subgraph in both modes, dynamic kernels are slower than their static counterparts.

\textbf{Vectorization path selection.} The dynamic-shape kernel generator can guard only a single vector length at runtime, but TorchVision models require different optimal widths, so the vectorised path often does not trigger. Adding runtime checks for multiple widths (1, 2, 4, 8) would add overhead; vectorisation typically improves single-kernel performance by 20--25\%.

\textbf{Layout propagation.} For convolutional networks, layout propagation is an important optimisation. PyTorch's default layout is row-major (NCHW), but convolution kernels are generally faster with NHWC layout because it better leverages tensor cores. Deep-learning compilers typically insert layout conversions at the network boundaries and keep intermediate tensors in NHWC. In dynamic mode all dimensions are unknown at compile time, so we cannot determine whether a tensor is NCHW or NHWC and cannot apply layout optimisation.

\textbf{Future work.} Several improvements could alleviate these limitations. In practice not all dimensions are unknown---if most are statically known and only a few are dynamic, we could constant-fold more expressions during kernel generation. We could also build layout expressions parameterised by shape dimension variables, deriving the layout (NCHW vs.\ NHWC) at compile time and re-enabling layout optimisation.

\subsection{Compilation cost}
\label{sec:eval-compile}


Figure~\ref{fig:compile-time} compares compilation time for \dlcb{}-static and \dlcb{}-dynamic
across the TorchVision suite.
Both \dlcb{}-static and \dlcb{}-dynamic compile significantly faster than \texttt{torch.compile},
with a median compile time of 0.84\,s versus 9.4\,s for \texttt{torch.compile} — roughly a
$10\times$ reduction (Figure~\ref{fig:compile-time}). Crucially, the \dlcb{}-dynamic compilation cost is an \emph{ahead-of-time}
overhead incurred once at model load; unlike \texttt{torch.compile}, no recompilation is triggered
at runtime when new shapes are encountered.

\section{Related Work}

\paragraph{Deep learning compilers with static shapes.}

The dominant approach in production deep-learning compilation requires fully static tensor shapes
at compile time.  XLA~\cite{xla} compiles HLO graphs with concrete dimensions and recompiles from
scratch on every shape change.  TVM~\cite{chen2018tvm} adds a learned cost-model autotuner whose
search is tied to a specific shape.  Halide~\cite{ragan2013halide} and Triton~\cite{tillet2019triton}
expose tiled GPU kernels but leave tile sizes and dimensions as compile-time constants.
Glow~\cite{rotem2018glow} and Relay~\cite{roesch2019relay} likewise assume concrete shapes during
code generation.  MLIR~\cite{lattner2021mlir} represents symbolic dimensions natively and provides
a shape dialect for constraint propagation; \dlcb{}'s constraint language is a lightweight
CUDA-targeted analogue.  Stripe~\cite{zerrell2019stripe} expresses tiling schedules parametrically
over a polyhedral model, but keeps dimension sizes as compile-time constants.

\paragraph{Dynamic shape support.}

Several recent systems attack dynamic shapes at the graph level.  DISC~\cite{zhu2021disc}
(Alibaba BladeDISC), the closest published architecture to \dlcb{}, separates a CPU
shape-inference sub-graph from GPU computation so that shapes are evaluated at runtime without
recompilation.  \dlcb{} operates at the kernel-parameter level rather than graph level, using a
constraint solver to classify dimensions at compile time.  Nimble~\cite{shen2021nimble} extends
TVM with a dynamic type system and a VM dispatching to per-shape kernels with JIT fallback;
\dlcb{}'s AOT path avoids runtime recompilation entirely for known rank.
TSCompiler~\cite{tscompiler2024} partitions operators by symbolic shape dependencies for
graph-level scheduling.  MikPoly~\cite{cui2024mikpoly} assembles micro-kernels on-the-fly to cover
dynamic shapes by a different mechanism from \dlcb{}'s parameterised kernels.
Brainstorm~\cite{cui2023brainstorm} targets dynamic control flow rather than dimensions.

\paragraph{JIT compilation and the PyTorch ecosystem.}

PyTorch~2~\cite{ansel2024pytorch2} (TorchDynamo + TorchInductor + SymInt) defines the state of the
art that \dlcb{} either replaces on the AOT path or falls back to on the JIT path.
nvFuser~\cite{nvfuser}, \dlcb{}'s primary performance baseline, generates fused CUDA kernels at
runtime but pays compilation latency on every distinct shape.  Ansor~\cite{zheng2020ansor} and
AutoTVM~\cite{chen2018autotvm} automate TVM schedule search with learned cost models but require
re-search per shape, motivating shape-agnostic AOT alternatives.  Proteus~\cite{proteus2025}
re-specialises existing GPU kernels via JIT at runtime, bounding the compile/runtime tradeoff from
the opposite direction to \dlcb{}.

\paragraph{Shape inference and type systems.}

Hattori et al.~\cite{hattori2020semistatic} combine dynamic execution tracing with static symbolic
analysis to classify tensor dimensions as concrete or symbolic, directly paralleling \dlcb{}'s
constraint-based static and dynamic classification.  ShapeIt~\cite{shapeit2024} instruments Python
bytecode to infer likely symbolic equalities between tensor dimensions at runtime, producing
annotations that could feed a compiler such as \dlcb{}.  ONNX shape inference~\cite{onnx} defines
per-operator rules distinguishing static dimensions, symbolic \texttt{dim\_param} values, and
unknown dimensions, providing an industrial precedent for constraint-propagating dimension
labelling.

\paragraph{Polyhedral compilation and parameterised kernels.}

PLUTO~\cite{bondhugula2008pluto} is the seminal framework for automatic affine tiling and
parallelisation; it generates code with tile sizes as compile-time constants, whereas \dlcb{} defers
those parameters to a lightweight host computation at launch time.  PPCG~\cite{verdoolaege2013ppcg}
generates multi-level tiled CUDA kernels from polyhedral loop nests, passing problem size $N$ as a
runtime argument---the same pattern \dlcb{} generalises to the full set of dynamic tensor dimensions.
Tiramisu~\cite{baghdadi2019tiramisu} expresses schedule independently of concrete problem sizes
across CPUs, GPUs, and distributed systems, demonstrating that AOT-compiled polyhedral schedules can
be parameterised by runtime values.

\section{Conclusion}

We have presented \dlcb{}, a deep learning compiler that resolves the tension
between ahead-of-time compilation and dynamic tensor shapes through a
constraint-based approach.  For the common case in which tensor rank is
statically known, \dlcb{} generates CUDA kernels once, ahead of time, that
execute correctly across a range of shapes by receiving dynamic dimensions as
arguments at launch time.  A constraint solver partitions dimensions into
compile-time constants and runtime parameters, and compiles residual
constraints into a compact host program that runs on the CPU before each
kernel launch to verify shape invariants and supply the required arguments.
When AOT compilation is not possible---because rank itself is unknown---the
system falls back transparently to JIT compilation, providing a unified
interface that always produces a runnable artifact.

A key practical contribution is the automatic shape generalization pass in
pytorch-dlcb.  Users need not annotate their programs with symbolic
dimensions: the pass identifies which dimensions are safe to abstract within
a statically traced PyTorch program and generates the corresponding
parameterised kernels automatically, lowering the barrier to deploying
AOT-compiled kernels in production settings where input shapes vary.

Our evaluation shows that \dlcb{}'s dynamic-shape kernels approach
fully-static compilation performance, demonstrating that the overhead of
parameterising kernels by runtime dimensions is small in practice.  A single
dynamic binary also eliminates the repeated compilation cost that static
approaches incur as input shapes vary across a deployment lifetime.

\bibliographystyle{ACM-Reference-Format}
\bibliography{refs}

\end{document}